\documentclass[aps,prb,preprint,superscriptaddress]{revtex4-1}
\usepackage{macros_BW}

\begin{document}

\title{Full elasticity tensor from thermal diffuse scattering}

\author{Bj\"orn Wehinger}
\email[]{bjorn.wehinger@unige.ch}
\affiliation{Department of Quantum Matter Physics, University of Geneva, 24, Quai Ernest Ansermet, CH-1211 Gen\`eve, Switzerland}
\affiliation{Laboratory for Neutron Scattering and Imaging, Paul Scherrer Institute, CH-5232
Villigen PSI, Switzerland}

\author{Alessandro Mirone}
\email[]{mirone@esrf.fr}
\affiliation{European Synchrotron Radiation Facility, 71, Avenue des Martyrs, F-38000 Grenoble, France}

\author{Michael Krisch}
\affiliation{European Synchrotron Radiation Facility, 71, Avenue des Martyrs, F-38000 Grenoble, France}

\author{Alexe\"i Bosak}
\affiliation{European Synchrotron Radiation Facility, 71, Avenue des Martyrs, F-38000 Grenoble, France}

\date{\today}

\begin{abstract}
We present a method for the precise determination of the full elasticity tensor from a single crystal diffraction experiment using monochromatic X-rays. For the two benchmark systems calcite and magnesium oxide we show that the measurement of thermal diffuse scattering in the proximity of Bragg reflections provides accurate values of the complete set of elastic constants. This approach allows for a reliable and model free determination of the elastic properties and can be performed together with crystal structure investigation in the same experiment. 

\end{abstract}

\maketitle

\section{Introduction}
The elasticity tensor is the fundamental quantity for describing elastic waves and determines eventual anisotropic compression and sound velocities in crystalline materials \cite{fedorov_PP_1968}. The elastic properties define mechanical properties of materials and influence phase stability \cite{cowley_prb_1976}. Accurate measurements of the elasticity are of principal interest for the description of seismological waves and their discontinuities, which allow for decisive conclusions on composition, temperature and pressure of Earth's interior \cite{birch_jgr_1952}. In condensed matter physics elastic constants are important in the study of quantum phase transitions in systems with pronounced interaction of phonons with other quasi-particles. Such interaction may include electron-phonon coupling with applicability to superconductors \cite{bolef_pr_1963,boemmel_pr_1955,migliori_prl_1990} as well as spin-phonon coupling with interesting anomalies in low dimensional spin systems \cite{almond_pla_1975,wolf_epl_1999}. 
The most commonly used experimental techniques to determine the coefficients of the elasticity tensor -- the so-called elastic constants -- are ultrasound measurements or Brillouin scattering. Ultrasound measurements are limited to relatively large crystals with well-defined faces, measurements at extreme conditions such as high pressures or high magnetic fields are very challenging. Brillouin scattering can be performed at high pressures but is difficult for opaque materials. Alternatively, the elasticity tensor can be extracted from inelastic x-ray or neutron scattering, which becomes efficient for crystals of low symmetry if combined with calculations from first principles \cite{wehinger_grl_2016}. Inelastic x-ray scattering can be performed at extreme pressures using diamond anvil cells and can be combined with low or high temperatures \cite{antonangeli_prl_2004,antonangeli_epsl_2012}.  

A complete formalism to derive ratios of elastic constants from thermal diffuse scattering (TDS) measured with energy integrating detectors has been derived in the 1960ies \cite{wooster_OUP_1962}. Historically, the thermal nature of diffuse scattering was noticed already in the 1920ies \cite{faxen_zp_1923, waller_zp_1923} and the first phonon dispersion relation were in fact determined from TDS of X-rays \cite{olmer_ac_1948}. The currently available high flux and brilliant x-ray beams from synchrotrons in combination with bi-dimensional single photon counting x-ray detectors with good quantum efficiency and no readout noise attracted new interest in TDS studies \cite{holt_prl_1999, xu_zkri_2005, bosak_jap_2015}. 
By the use of force constant models it is possible to determine the phonon dispersion relations, while a model-free reconstruction of the lattice dynamics can be realized only for mono-atomic crystals \cite{bosak_aca_2008}. 

In this Letter, we present a method for the precise determination of the full elasticity tensor from a single crystal diffraction experiment using a model free data analysis for arbitrary crystal symmetries. For the two benchmark systems calcite and magnesium oxide we show that the measurement of thermal diffuse scattering at two close temperatures is sufficient to obtain the full elasticity tensor in absolute units within remarkable accuracy. We discuss the influence of additional contributions to the diffuse scattering and evaluate multiple phonon scattering with help of first principle calculations using density functional perturbation theory.

\section{Experimental details}
Diffuse x-ray scattering intensities were collected in transmission geometry using a single photon counting pixel detector with no readout noise and large dynamical range. High quality single crystals were prepared by mechanical cutting and polishing; surface defects were removed by gentle etching. Temperature was controlled by a nitrogen cryostream.
The experiment on calcite was conducted on ID29 at the European Synchrotron Radiation Facility (ESRF) \cite{deSanctis_jsr_2012}. The sample was a rectangular bar with dimensions of approximately 50\, $\mu$m $\times$ 50\, $\mu$m $\times$ 1.5\, mm. Measurements were taken at 170\,K. Monocromatic X-rays with a wavelength of $\lambda=0.6968$\,\AA\,and a microfocus of 60\,$\mu$m $\times$ 30\,$\mu$m were used. The sample was rotated from $0$ to $100^{\circ}$ orthogonal to the beam with angular steps of 0.1$^{\circ}$. Scattering intensities were collected with a PILATUS 6M detector (Dectris, Baden, Switzerland), equipped with 300\,$\mu$m thick Si pixels of size 172$^2$\,$\mu$m$^2$, at a sample-detector distance of 300\,mm.
Measurements on magnesium oxide were performed at the Swiss Norwegian Beamline BM01A at the ESRF where we used a cubic single crystal with edge length of 2\,mm. The sample was measured at  $\lambda=0.68894$\,\AA\, and a temperature of 90 and 120\,K. The crystal was rotated by 360$^{\circ}$ with the same angular step as for calcite. Scattering intensities were recorded with a PILATUS 2M detector at a distance of 244\, mm from the sample. 

\section{Formalism}
\label{metodo}
Assuming the validity of both, adiabatic and harmonic approximations, the intensity of x-ray scattering from phonons for single and two phonon processes is given by

\begin{equation}
\label{eq:I1}
I_1 (\bs{Q}) = \frac{\hbar N I_{inc}}{2} \sum_{\nu} \Omega_{\bs{q},\nu} \big| \sum_{a} \frac{f_a(\bs{Q})}{\sqrt{m_a}} e^{-W_{a,\bs{Q}}} (\bs{Q} \bs{e}_{\bs{Q},\nu,a}) e^{-i \bs{Q} \bs{\tau}_a} \big| ^2
\end{equation}
and

\begin{eqnarray}
\label{eq:I2}
I_2 (\bs{Q}) & = & \frac{\hbar^2 N V I_{inc}}{8} \int \frac{d^3 \bs{q}}{(2\pi)^3} \sum_{\nu,\nu'} \Omega_{\bs{q},\nu} \Omega_{\bs{Q} - \bs{q},\nu'} \times \nonumber \\
& & \big| \sum_a \frac{f_a(\bs{Q})}{m_a} e^{-W_{a,\bs{Q}}} e^{-i \bs{Q} \bs{\tau}_a} (\bs{Q} \bs{e}_{\bs{q},\nu,a}) (\bs{Q} \bs{e}_{\bs{Q}- \bs{q},\nu,a}) \big|^2 \nonumber \\
& & 
\end{eqnarray}
with

\begin{equation}
\Omega_{\bs{q},\nu} = \frac{1}{\omega_{\bs{q},\nu}} coth \Big( \frac{\hbar \omega_{\bs{q},\nu}}{2 k_B T} \Big),
\end{equation}
respectively; see Refs. \cite{xu_zkri_2005,bosak_jap_2015} for a description in modern notation. Here, 
$N$ is the number of unit cells, $I_{inc}$ the incident beam intensity, $f$ the atomic scattering factor of atom $a$ with mass $m$ and (anisotropic) Debye Waller factor $W$. $\omega$ denotes the eigenfrequency and $\bs{e}$ the eigenvector of the phonon with wavevector $q$ (equal to the reduced momentum transfer) and branch $\nu$. $Q$ is the total scattering vector, $\tau$ the atomic basis vector within the unit cell and $V$ the unit cell volume, $T$ the temperature and $k_B$ the Boltzmann constant.
For small ${\bs{q}}$ the scattering intensities $I_1(\bs{Q}) \sim 1/\omega_{\bs q, \nu}^2$ and single phonon scattering is thus dominated by the contribution of the acoustic phonons.

Within the theory of elastic waves in crystals the equation of motion is given by 

\begin{equation}
\label{eq:el_motion}
\rho \omega^2 u_{i} = c_{ijlm} k_j k_l u_m,
\end{equation}
where $c_{ijlm}$ is the tensor of elastic constants, $\rho$ the mass density and $ \bs{k} = k \bs{n} $ and $ \omega $ are the wave vector and the frequency of the elastic waves, respectively \cite{fedorov_PP_1968}. 

For fitting the elastic constants to the experimental intensities in the vicinity of Bragg reflections we calculate the scattering intensities as the sum of the contributions from the three acoustic branches. We therefore solve the equation of motion (Eq. \ref{eq:el_motion}) for a given crystal symmetry and calculate the scattering intensities by summing over the three acoustic phonon branches in Eqs. \ref{eq:I1} and \ref{eq:I2}. The thus obtained intensities are  renormalized by a vector $\bs{g}(\bs{Q})$ that accounts for absorption, polarisation and geometrical corrections. A vector $\bs{b}(\bs{Q})$ is added for the background. 

Given a set of experimental intensities $I^{exp}_{{\bs Q}, T} $ measured at $T$ over a set of reciprocal space points  $ \bs{Q} \in \left\{ Q_{exp}   \right\}$ in the proximity to Bragg reflections, we find the elasticity tensor $c$ by solving the optimization problem

\begin{equation}
\label{eq:monotemp}
c, {\bs{b}},{\bs{g}} =  \underset{c^\prime , {\bs{b}^\prime},{\bs{g}^\prime}}{\operatorname{argmin}} \left( \sum_{ {\bs{Q}}}  \left( I^{calc}_{{\bs{Q}}, T}( c^\prime, {\bs{b}^\prime} ,{\bs{g}^\prime}) -  I^{exp}_{{\bs{Q}}, T} \right)^2\right), 
\end{equation}  
where $I^{calc}_{{\bs{Q}}, T}$ is the calculated intensity containing contributions from one-phonon and eventually $(n>1)$-phonon processes.
$c$, $\bs{b}$ and $\bs{g}$ are the fit parameters. In order to reduce the free parameters $c$ is constrained to the crystal symmetry. $\bs{b}$ and $\bs{g}$ are kept constant in the vicinity of individual Bragg reflections. Such approximation for $\bs{b}$ is justified by the fact that diffuse scattering due to additional contributions varies much less across reciprocal space than TDS for small $\bs{q}$. The variation of corrections for absorption, polarisation and planar projection is small for the employed scattering geometry at small $\bs{q}$ and thus justifies the approximation for $\bs{g}$.

Solving the minimization problem Eq. \ref{eq:monotemp} requires that the diffuse scattering is due to phonon scattering only. It can provide absolute values of the elastic constants if absolute intensities are known. If not, the elastic tensor is determined upon a single scaling factor and the absolute values of the elastic constants can be obtained if constraint to the adiabatic bulk modulus via the Reuss-Voigt-Hill relation.

Another option consists of measuring scattering intensities at slightly different temperatures. In fact, diffuse scattering from static disorder, air scattering, and fluorescence displays a much smaller temperature dependence than TDS, and can therefore be isolated in a difference measurement. 

For experimental data at two slightly different temperatures $T_1$ and $T_2$ measured in the same geometry we can solve the minimization problem
\begin{equation}
\label{eq:multitemp}
  \begin{split}
  c, {\bs{b}_2} , {\bs{b}_1}, {\bs{g}} =  \underset{c^\prime , {\bs{b}_2^\prime} , {\bs{b}_1^{\prime}} , {\bs{g}^\prime}}{\operatorname{argmin}} \left( \sum_{ {\bs{Q}}}
  \Bigg(  \Big(
  I^{calc}_{{\bs{Q}}, T_2}( c^\prime, {\bs{b}_2^\prime}  ,{\bs{g}^\prime})-
  I^{calc}_{{\bs{Q}}, T_1}( c^\prime, {\bs{b}_1^{\prime}} ,{\bs{g}^\prime})
  \Big)
  -
  \right. \\ \left.
  \Big(
  I^{exp}_{{\bs{Q}}, T_2}   - I^{exp}_{{\bs{Q}}, T_1} \Big)
  \Bigg)^2
  \right),
  \end{split}
\end{equation}
which we call multi-temperature method. It neglects the variation of the elastic constants over the temperature interval $T_1$,$T_2$. The intensity difference must be compared to the variations of the Bose factor $coth\left( \frac{\hbar \omega_{{\bs{q}},\nu} }{2kT}  \right)$, which is approximately proportional to the temperature 
at high temperatures ($\hbar \omega >  k T$). This implies that, at constant $c$, the intensities measured at  $T_1$ and $T_2$ are almost proportional to each other if one-phonon processes are predominant. Therefore, at such high temperatures, only ratios of elastic constants can be obtained if the scattering intensities are unknown on an absolute scale. 
At lower temperatures the acoustic branches span two regions, a low frequency one with $\hbar \omega \ll k T$ and one with higher frequencies $\hbar \omega >  k T$. In such case the intensities of the two measurements become linearly independent. 
The knowledge of $T$ determines the absolute scale to which $\omega$ compares, and, as a consequence, absolute values of the elastic constants can be obtained even for unknown absolute intensities. The temperature must be low enough so that the condition $\hbar \omega > k T$ is realized in regions of ${\bs{q}}$ where the elastic approximation is fulfilled.

\section{Lattice dynamics calculations}
Lattice dynamics calculations were carried out from first principles employing density functional perturbation theory \cite{gonze_prb_1997_2} as implemented in the CASTEP code \cite{clark_zkri_2005,refson_prb_2006}. 
For both, calcite and magnesium oxide we used the local density approximation within a plane wave basis set and pseudo-potentials of the optimized form \cite{rappe_prb_1990}. The sampling of the electronic structure and the plane wave cut-off energy were chosen to ensure the convergence of internal forces to $<$ 1e$^{-3}$\, eV/\AA. The acoustic sum rule was enforced to ensure translational symmetry. Eigenfrequencies and eigenvectors of the acoustic branches were replaced in order to reproduce the elastic approximation close to $\Gamma$. Fourier interpolation was employed for the computation of dynamical matrices at arbitrary $\bs{q}$. The calculated dynamical matrices were used to compute the Debye Waller factors and scattering intensities were calculated via Eqs. \ref{eq:I1} and \ref{eq:I2}. The calculations are used for the correct selection of $\bs{q}$-values to be fitted and for the evaluation of multiple phonon scattering and contribution of optical phonons. 

\section{Results}
\begin{figure}[tbp]
\vskip-2.0cm
  \begin{tikzpicture}[scale=0.8]
    \node[anchor=south west,inner sep=0] at (0,0) {\includegraphics[width=0.75\textwidth]{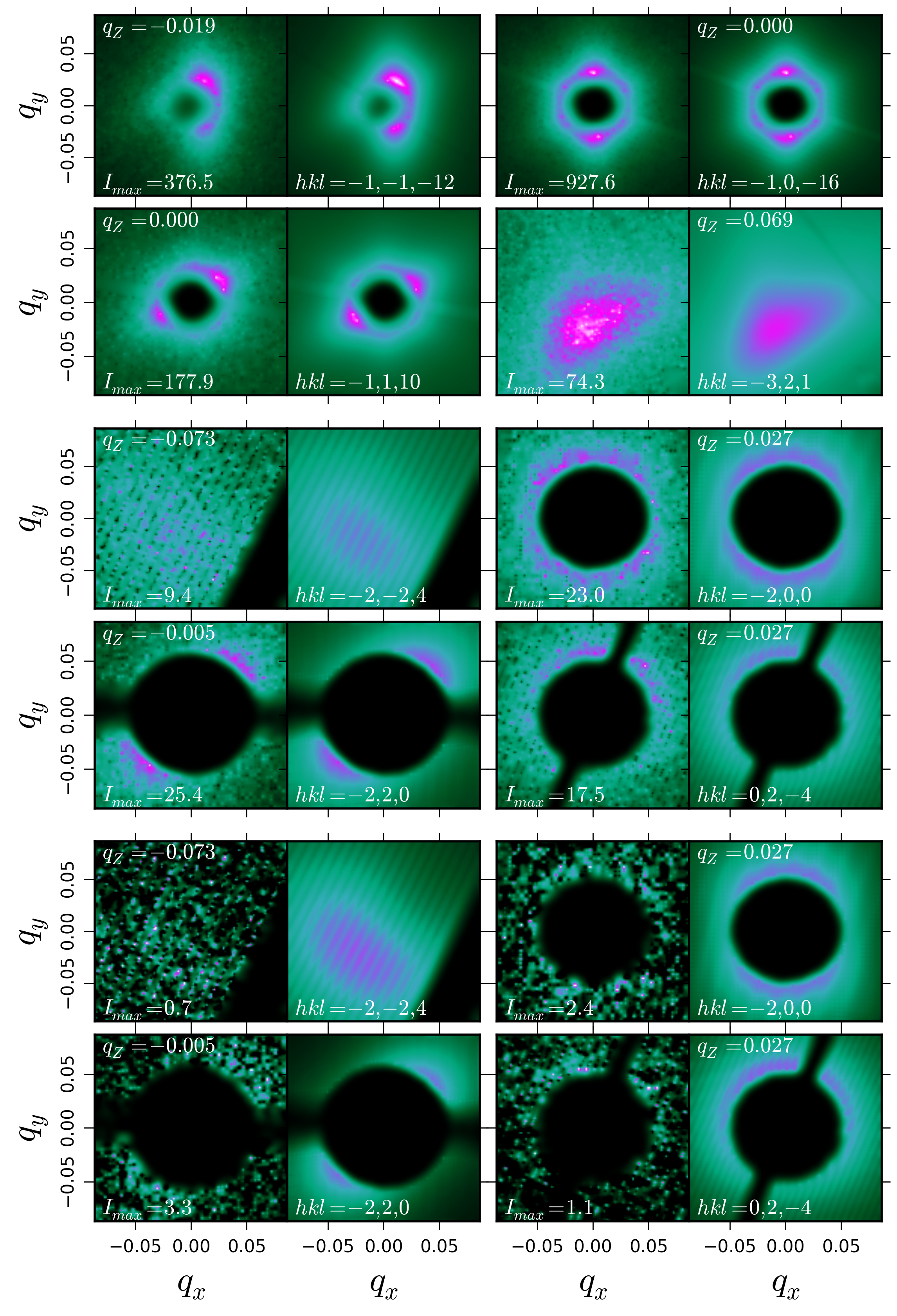}};
    \node[ draw,fill=white,draw,align=center] at (8.1,22.5) {Calcite, $T=170$K};
    \node[ draw,fill=white,align=center] at (8.1,15.3) {MgO, $T=90$K};
    \node[black,draw, align=center,fill=white] at (8.1,8.2) {MgO, $I(120\text{K})-I(90\text{K})$};
\end{tikzpicture}
  \caption{Graphical rendering of measured diffuse scattering and calculated TDS. Considered cases are: Calcite at $T$ = 170\,K for a ROI of  $q \in [0.03,0.15]$ (top panels),
MgO with $q \in [0.06,0.15]$ at $T$ = 90\,K (middle panels) and with the multi-temperature method at $T$ = 90\,K, 120\,K (bottom panels). Each image shows a cross section of the reciprocal space, in the neighborhood of the select Bragg peak, and for a given $q_z = \left| \Delta {\mathbf Q}_z \right| /\left| a^* \right|$. The data are grouped by pairs, with the experimental and calculated intensity distribution on the left and right side, respectively. The scattering intensity is shown on a linear color scale from black (zero) to white (given maximal intensity $I_{max}$). The reverse interpolation, which is only used for graphical rendering, results in some artifacts, for example some intensity remains outside the ROI. The stripes visible at $hkl = \bar{2}, \bar{2}, 4$ and $0,\bar{2},4$ are due to the finite angular steps in the measurement and each stripe corresponds to the contribution of a single diffraction pattern.}
  \label{fig:TDS}
\end{figure}
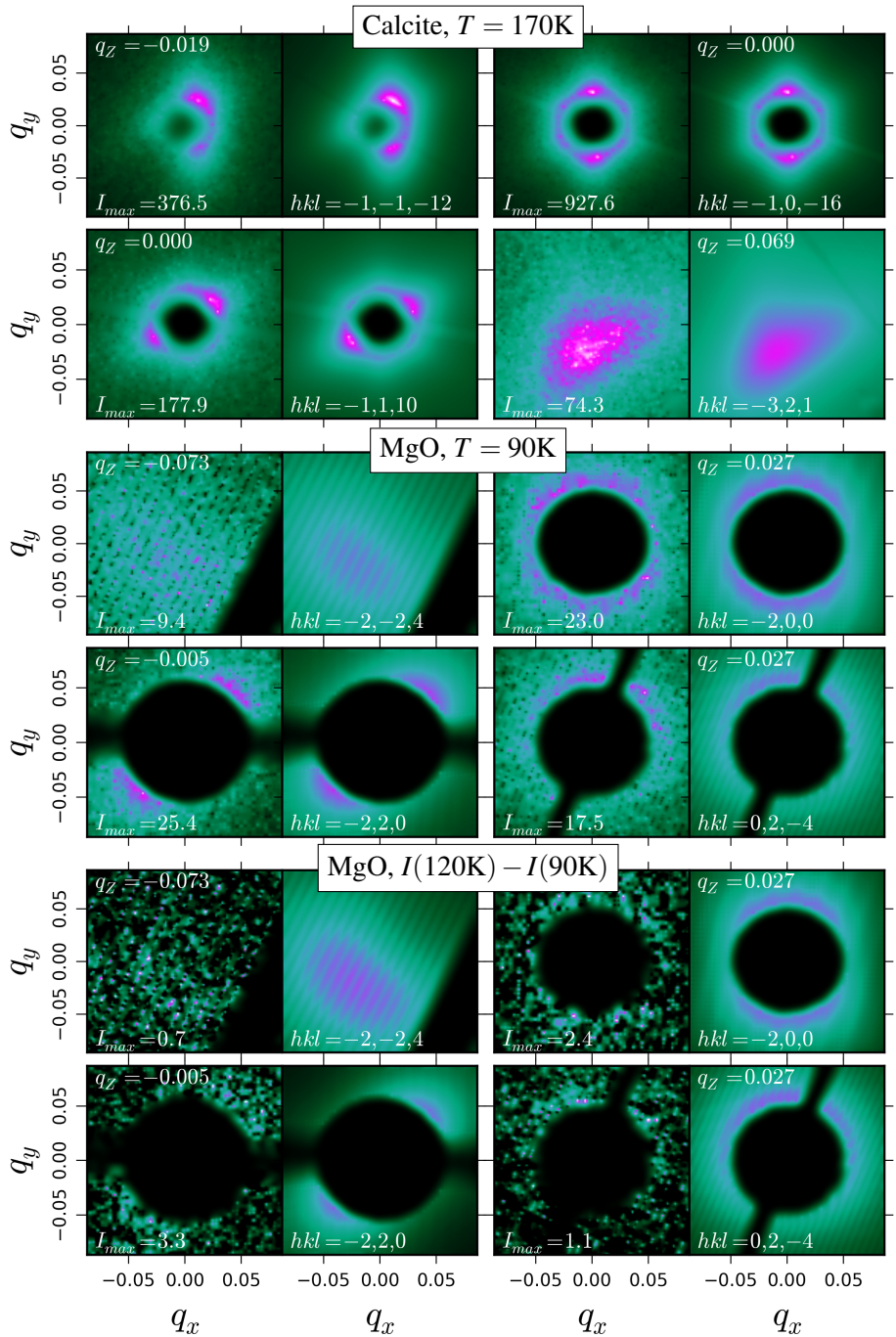

\subsection{Calcite at 170 Kelvin}

\begin{ruledtabular}
\begin{table}[tp]
  \caption{Elastic constants of calcite at 170\, K. The values derived from fitting TDS are obtained for $q \in [0.06,0.15]$ and normalized by the adiabatic bulk modulus. Experimental reference values from ultrasound measurements (Ref. \cite{dandekar_jap_1968}) and the relative difference between the values derives from TDS and ultrasound are reported for comparison. Theoretical values were derived from fitting calculated TDS intensities within the same ROI ($q \in [0.06,0.15]$) to the contribution of (i) acoustic branches only, (ii) all phonon branches  and (iii) acoustic branches including two-phonon scattering.}
  {
    \setlength\tabcolsep{1.5pt} 
    \begin{tabular}{lllllll}
      \hspace{20pt} &  TDS \hspace{5pt} & Ref. \cite{dandekar_jap_1968} \hspace{2pt} & rel. diff. & Calc. (i) & Calc. (ii) & Calc. (iii) \\
      \midrule
      $c_{11}$  & 156  & 155.1 & 0.6\%  & 154.3 & 153.5 & 154.2  \\
      $c_{13}$  & 57.0 & 57.69 & 1.1\%  & 57.74 & 57.86 & 57.81  \\
      $c_{15}$  & 21.2 & 21.51 & 1.6\%  & 21.11 & 20.86 & 20.94  \\
      $c_{33}$  & 87.6 & 87.12 & 0.6\%  & 86.91 & 86.88 & 86.80  \\
      $c_{44}$  & 35.7 & 34.42 & 3.7\%  & 33.95 & 33.60 & 33.75  \\
      $c_{66}$  & 48.0 & 47.80 & 0.4\%  & 46.53 & 46.06 & 46.36  \\
    \end{tabular} 
  }
   \label{table:calcite}
\end{table}
\end{ruledtabular}

The determination of the elastic constants from TDS measured at a single temperature is demonstrated for a small calcite single crystal. For the fit we consider the diffuse scattering intensity in the proximity of the $48$ most intense Bragg reflections. The regions of interest (ROI) were parametrized by the absolute value of the reduced wave vector $q$ in Cartesian coordinates with $q_i = \left| \Delta {\mathbf Q}_i \right| /\left| a^* \right|$ and the data points selected according to the criteria $ q \in [q_{min},q_{max}] $. The regions below $q_{min}$ are excluded to minimize the contribution from elastic scattering while the regions above $q_{max}$ are excluded in order to ensure the validity of the elastic approximation. 

Experimental scattering intensities and calculated TDS from the fitted elastic constants in the elastic approximation for first order scattering are shown for selected regions of reciprocal space in Fig. \ref{fig:TDS}. 
The plotted values are obtained from the irregular experimental data grid by inverting the interpolation equation. We do this using a few iterative steps and a Thikonov regularisation term.
Diffraction patterns that contain Bragg peaks were removed from the data treatment, because such images are affected by secondary scattering effects \cite{ramsteiner_jac_2009}. 

The elastic constants as obtained from fitting intensities of an ensemble of individual pixels are reported in Table \ref{table:calcite}. Eq. \ref{eq:monotemp} was used to determine $c$ upon a single scaling factor and the absolute values were obtained by normalization to the known adiabatic bulk modulus $K$. Remarkable agreement with literature values determined from ultrasound measurements \cite{dandekar_jap_1968} are obtained for a ROI $ q \in [0.06,0.15]$, which corresponds to approximately $2.1\times 10^7$ intensity points. The difference is in the order of 1\,\% for all elastic constants with exception of $c_{44}$.
The higher limit $q_{max}$ of the ROI is verified by \textit{ab initio} calculations. We therefore compute the scattering intensities from the calculated dynamical matrices and fit $c$ to it for different choices of ROI and compare the result to the expected value for $ \lim \bs{q} \rightarrow 0$.
The lower limit $q_{min}$ of the ROI must be chosen carefully, because very small momentum transfers might be affected by elastic scattering, as discussed further below. 
The contribution of optical phonons and multiple phonon scattering is evaluated by computing $I_1 (\bs{Q})$ and $I_2 (\bs{Q})$ including all phonon branches. A fit of elastic constants to these computed scattering intensities for the same ROI results in a maximal relative deviation of 1.2\,\% if optical phonons are considered and only 0.8\,\% maximal relative deviation if second order phonon scattering is included (see Table \ref{table:calcite}).

\subsection{Magnesium oxide measured at two temperatures}

\begin{ruledtabular}
\begin{table}[tp]
     \caption{Elastic constants of MgO measured at $T_1$ = 90\, K and $T_2$ = 120\,K. The elastic constants were fitted for $q \in [0.06,0.15]$ applying the multi-temperature method for the two temperatures ($I(T_2 -T_1)$) and for $T_{1}$ only (Fit $I(T_1)$, values re-scaled to the reference value of c$_{11}$).  Ultrasound data \cite{marklund_ps_1971} for the two temperatures and the relative difference between the average reference values and the results of the multi-temperature method are listed for comparison.} 
  {
    \setlength\tabcolsep{5pt} 
    \begin{tabular}{llllll}
      & Fit $I(T_2) - I(T_1)$ & Fit $I(T_1)$ & Ref. \cite{marklund_ps_1971} & Ref. \cite{marklund_ps_1971} \\
      & & c$_{11}$ rescaled  & $T$ = 90\, K & $T$ = 120\, K & rel. diff. \\
      \midrule
      $c_{11}$  & 300 & 306 & 306.1  & 305.4 & 2.0\% \\
      $c_{44}$  & 151 &194  & 157.2  & 156.9 & 3.7\% \\
      $c_{12}$  & 89   & 39   & 94.07  & 94.26 & 5.7\% \\
    \end{tabular} 
  }
     \label{table:mgo}
\end{table}
\end{ruledtabular}

The diffuse scattering in MgO at $T$ = 90\, K is much less structured than the one of calcite due to the higher cubic crystal symmetry, see Fig. \ref{fig:TDS}. For fitting the elastic constants both methods, Eq. \ref{eq:monotemp} and \ref{eq:multitemp}, were employed considering diffuse scattering in the proximity of $78$ of the  most intense Bragg reflections.
The results are reported in Table \ref{table:mgo}. Fitting a single temperature with Eq. \ref{eq:monotemp} is performed to obtain $c$ upon a uniform scaling factor. Due to the large discrepancy in $c_{12}$ the values are scaled to the experimental value of $c_{11}$ from ultrasound measurements instead of applying a scaling to the bulk modulus. The results are rather unsatisfactory compared to literature data \cite{marklund_ps_1971}. This demonstrates the influence of additional diffuse scattering due to elastic scattering. We thus employ the multi-temperature method (Eq. \ref{eq:multitemp}) and fit the intensity differences of diffuse scattering measured at two close temperatures, 90 and 120\, K, see Fig. \ref{fig:TDS}. The data seems noisy, but fitting the ensemble of approximately $1.7 \times 10^{7}$ pixels is sufficient for a well converged result. Using this strategy we obtain accurate values of the full elasticity tensor in absolute units, presented in Table \ref{table:mgo}. 

\section{Discussion}
The results on calcite and magnesium oxide show that the full elasticity tensor can be measured with high accuracy by a rigorous data treatment of diffuse  scattering in the proximity of Bragg reflections. The availability of single photon counting detectors with no readout noise together with a large number of independent intensity points are crucial for successful experiments. For single crystals of extreme high quality such as our investigated calcite crystal the measurement of TDS at a single temperature can be sufficient to obtain the correct ratio of all elastic constants. The absolute values may then be obtained by a normalization to the adiabatic bulk modulus. The optimized choice of the  ROI to be fitted is very important to ensure the validity of the elastic approximation and allows minimizing the contribution of elastic scattering. The best ROI can be found by a successive adjustments of $q_{min}$ and $q_{max}$ and is verified here with help of lattice dynamics calculations. Scattering contributions that vary slowly in reciprocal space can be taken into account in good approximation by a constant background to the diffuse scattering in the vicinity of individual Bragg reflections. Such contributions may include air scattering, fluorescence and Compton scattering. This includes as well the contribution of optical phonons and higher order phonon scattering as shown by evaluating calculated dynamical matrices. A de-convolution procedure for air scattering and beam shape may be envisaged in order to extend the ROI to smaller $q$ values, but this generally increases the noise level of the data. Secondary scattering effects like Bragg-diffuse scattering, where the Bragg reflected beam acts as a source of secondary diffuse scattering, are more difficult to treat and its contribution and scattering conditions are discussed elsewhere \cite{ramsteiner_jac_2009}. Here, we exclude all diffraction patterns that may be affected by such effects. 
If the diffuse scattering close to Bragg reflections is affected by elastic scattering the multi-temperature method might be a good choice. This method is based on the fact that TDS has a much stronger temperature dependence than other sources of diffuse scattering. Elastic and quasi-elastic contributions can thus be subtracted to a good approximation. Absolute values of the elasticity tensor can finally be obtained if the temperature interval is chosen such that the intensities become linear independent. This is shown for MgO. In addition to TDS we observe elastic diffuse scattering which we attribute to crystal defects. A careful measurement of absolute intensities may be a promising alternative to extract absolute values of the elastic constants but is likely less practical. 
In this study we compute the Debye Waller factors from first principles. However, they can also be obtained from experiment using x-ray diffraction employing the same scattering geometry.

\section{Conclusion and outlook}
In summary, we have shown that accurate values of the full elasticity tensor can be obtained from a simple diffraction experiment on single crystals. Our method opens the perspective to determine elastic properties together with crystal structure investigations and thus under the same experimental conditions. This implies a broad applicability in material science, geophysics and in the investigation of sound wave anomalies due to fundamental interactions in condensed matter physics. The achieved accuracy can compare with the standard methods such as ultrasound measurements and Brillouin scattering with the advantage of applicability to very small and opaque crystals of arbitrary shape. The proposed methodology can be extended to measurements at extreme conditions such as high pressures, high or low temperatures or high magnetic fields. The contribution from the sample environment might be treated with the multi-temperature method together with a deconvolution procedure for the treatment of temperature independent contributions. TDS from diamond in high pressure cells, for example, might be modeled or measured and then separated from the data by deconvolution. The application to high pressures is particularly interesting for the establishment of absolute pressure scales in a single experiment. At very low temperatures the proposed strategy is expected to work well and potentially very useful in the study of spin-lattice coupling. At temperatures relevant for geophysical processes the scattering intensities will not be linear independent but absolute values can be obtained if the adiabatic bulk modulus is known. 

\section{Acknowledgment}

We thank Dmitry Chernyshov, Daniele de Sanctis and Harald Reichert for providing beam time and fruitful discussions on measurement and data treatment of diffuse scattering. Gael Goret is acknowledged for help in programming the first versions of the reciprocal space reconstruction routines.
This work was supported by the European Community's Seventh Framework Programme (FP7/2007-2013) under grant agreement no. 290605 (PSI-FELLOW/COFUND) and resources of the European Synchrotron Radiation Facility.

\bibliographystyle{apsrev4-1_BW}
\bibliography{tds2el_refs}

\end{document}